\documentstyle[11pt,IAUS212,twoside,epsf]{article}

\def\edcomment#1{\iffalse\marginpar{\raggedright\sl#1\/}\else\relax\fi}
\reversemarginpar

\begin{document}
\vspace*{1cm}
\title{The Wind Momentum-Luminosity relationship: blue supergiants in the
spiral galaxy NGC 300}
\author{Fabio Bresolin, Rolf-Peter Kudritzki}
\affil{Institute for Astronomy, 2680 Woodlawn Drive, 96822 Honolulu, USA}
\author{Wolfgang Gieren}
\affil{Universidad de Concepci\'{o}n, Departamento de Fisica, Casilla 160-C, Concepci\'{o}n, Chile}

\begin{abstract}
We have carried out the wind analysis of six A-type supergiants in NGC
300. The derived Wind Momentum-Luminosity Relationship is compared
with that of Galactic and M31 blue supergiants and with theoretical
models.
\end{abstract}

\section{Introduction}

The theory of radiatively driven winds predicts a relationship between
the modified wind momentum and a power of stellar luminosity:
$$\log{(\dot{M}\,v_\infty\,R)} = x\log{L} + const$$

\noindent
wher $\dot{M}$ is the mass-loss rate, $v_\infty$ the wind terminal
velocity and $R$ the stellar radius.  Empirical verifications of this
Wind Momentum-Luminosity Relationship (WLR) have so far been carried
out for O stars in the Milky Way and the Magellanic Clouds by Puls et
al.~(1996), and for Galactic and M31 B and A supergiants by Kudritzki
et al.~(1999). The brightest A-type supergiants in galaxies are
extremely bright, with $M_V$ up to $-10$, making the WLR a potential
extragalactic distance indicator. However, extensive work still needs
to be carried out on the empirical calibration of the relation and on
the theoretical modeling of the effects of stellar metallicity and
spectral type.

\section{A-type supergiants in NGC~300}
We have carried out a spectroscopic survey of blue supergiants in the
Sculptor galaxy NGC~300 (Cepheid distance $\sim$ 2 Mpc) with
multiobject spectroscopy at 5~\AA\/ resolution with VLT/FORS (Bresolin
et al.~2002). These data enable us to determine spectral type,
metallicity, effective temperature and reddening of the confirmed blue
supergiants. In addition, the stellar mass-loss rate can be derived
from the H$\alpha$ line profile.

For our analysis we have selected six late-B to early-A
(B8-A2) supergiants, whose spectra appear uncontaminated by nebular
emission. Stellar gravities and mass-loss rates have been measured by
means of profile fits to the H$\gamma$ and H$\alpha$ lines, adopting
the NLTE line formation code of Santolaya-Rey, Puls \& Herrero
(1997). The modified wind momentum has been derived from the measured
$\dot{M}$ and the known distance (which, together with the apparent
magnitude, provides the stellar radius), and by assuming a terminal
velocity $v_\infty=150$ km/s. This is currently the major source of
uncertainty, since $v_\infty$ cannot be measured at the resolution
of our spectra.

We show in Fig.~1 the resulting WLR for the six supergiants in NGC~300
(squares), compared with the relationship for an equal number of
A-type supergiants in the Milky Way and M31 (circles; from Kudritzki
et al.~1999). Also plotted are new theoretical models accounting for
the metallicity dependence of the wind momentum, calculated for solar
and 0.4 solar metallicity. The latter value corresponds to the oxygen
abundance derived from the H\,{\sc ii} region galactic radial abundance gradient
(Zaritsky, Kennicutt \& Huchra 1994) and measured at the mean
galactocentric distance of our supergiant sample ($\sim$ 5 Kpc).  We
find a reasonable agreement between the data points and the models,
which predict $\dot{M}\,v_\infty\,R\propto Z^\alpha$, with
$\alpha\simeq0.8$. Our next step will be to further investigate the
systematics of the WLR in terms of stellar parameters ($\log g$,
$T_{\em eff}$, metallicity).

\begin{figure}[ht]
\plotfiddle{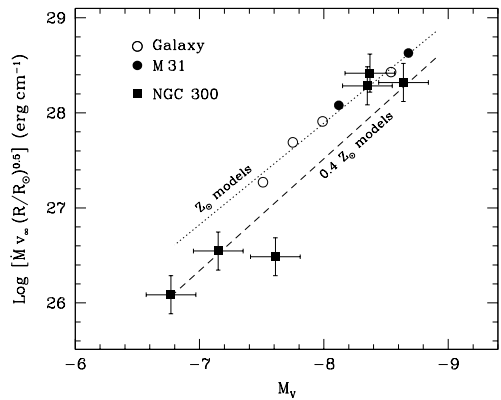}{5cm}{0}{35}{35}{-90}{-80}
\caption{The WLR of the six A-type supergiants analyzed in NGC~300, compared
with stars in the Galaxy and M31. Theoretical models at solar and 0.4
solar metallicity are also shown.}
\end{figure}


\begin{references}
\reference{Bresolin, F., Gieren, W., Kudritzki, R.-P., Pietrzy\'{n}ski, G.,
\& Przybilla, N., 2002, \apj, 567, 277}

\reference{Kudritzki, R.-P., Puls, J., Lennon, D.J., Venn, K.A.,
Reetz, J., Najarro, F., McCarthy, J.K., \& Herrero, A. 1999, \aap,
350, 970}

\reference{Puls, J., Kudritzki, R.-P., Herrero, A., et al. 1996, \aap,
305, 171}

\reference{Santolaya-Rey, A.E., Puls, J., \& Herrero, A. 1997, \aap,
323, 488}

\reference{Zaritsky, D., Kennicutt, R.C., \& Huchra, J.P. 1994, \apj,
420, 87}

\end{references}
\end{document}